\documentstyle [11pt,twoside]{article}
\input{./epsf.tex}
\def\la{\mathrel{\mathchoice {\vcenter{\offinterlineskip\halign{\hfil
$\reset@font\displaystyle##$\hfil\cr<\cr\sim\cr}}}
{\vcenter{\offinterlineskip\halign{\hfil$\reset@font\textstyle##$\hfil\cr
<\cr\sim\cr}}}
{\vcenter{\offinterlineskip\halign{\hfil$\reset@font\scriptstyle##$\hfil\cr
<\cr\sim\cr}}}
{\vcenter{\offinterlineskip\halign{\hfil$\reset@font\scriptscriptstyle##$\hfil\cr
<\cr\sim\cr}}}}}

\def\ga{\mathrel{\mathchoice {\vcenter{\offinterlineskip\halign{\hfil
$\reset@font\displaystyle##$\hfil\cr>\cr\sim\cr}}}
{\vcenter{\offinterlineskip\halign{\hfil$\reset@font\textstyle##$\hfil\cr
>\cr\sim\cr}}}
{\vcenter{\offinterlineskip\halign{\hfil$\reset@font\scriptstyle##$\hfil\cr
>\cr\sim\cr}}}
{\vcenter{\offinterlineskip\halign{\hfil$\reset@font\scriptscriptstyle##$\hfil\cr
>\cr\sim\cr}}}}}

\begin{document}
\centerline{\Large\bf Energy trapping in loaded string models with}
\vskip .1in
\centerline{\Large\bf long- and short-range couplings}
\vskip .2in
\centerline{\bf Ilya V. Pogorelov$^{1}$ and Henry E. Kandrup$^{1,2,3}$}
\vskip .25in
\centerline{$^{1}$ Department of Physics, University of Florida, Gainesville,
FL 32611}
\centerline{$^{2}$ Department of Astronomy, University of Florida, 
Gainesville,
FL 32611}
\centerline{$^{3}$ Institute for Fundamental Theory, University of Florida, 
Gainesville, FL 32611}
\vskip 0.5in
\begin{abstract}
\par\noindent
{\small This note illustrates the possibility in simple loaded string models
of trapping most of the system energy in a single degree of freedom for very
long times, demonstrating in particular that the robustness of the trapping 
is enhanced by increasing the `connectance' of the system, {\em i.e.,} the
extent to which many degrees of freedom are coupled directly by the 
interaction Hamiltonian, and/or the strength of the couplings.}
\end{abstract}
\par\noindent

The work described here was motivated by a desire to understand the 
qualitative difference in dynamical behavior in systems interacting
via short- and long-range couplings. 
In particular, how do bulk properties involving mixing and relaxation 
vary as the strength or the range of the interaction is increased?
Considerable work has been done on systems like the Fermi-Pasta-Ulam
({\em FPU}) model, which involve only nearest neighbor couplings 
(see, {\em e.g.,} Ford 1992). However, comparatively little is known about 
the phenomenology of systems which manifest longer range couplings. How
does the dynamics change for high-connectance systems (see, {\em e.g.,}
Froeschl\'e 1978), which couple directly all, or almost all, the degrees
of freedom?

When considering systems with longer range couplings, even the natural
language in which to describe the dynamics changes. Most work on 
{\em FPU}-type models has involved an analysis formulated in terms of the 
modes of the system. However, for systems with longer range interactions
it often becomes less obvious how to identify a natural set of modes.
More natural, perhaps, is to consider individual degrees of freedom
as fundamental, and to interpret the observed behavior in
terms of those degrees of freedom.

The work described here can be viewed as a prolegomenon towards more
systematic work underway which aims to study mixing properties in
self-consistent systems described in the continuum limit by partial
differential equations, such as the Vlasov-Poisson system of galactic
astronomy, plasma physics, or charged-particle beams.

As a simple example, consider a loaded string-type model consisting of
$N=16$ identical nonlinear oscillators arranged in a one-dimensional closed 
chain with dynamics generated by the Hamiltonian

$$H = \sum_{i=0}^{N-1} {\Bigl(} {1\over 2}p_{i}^{2} +{1\over 2}q_{i}^{2}
      +{1\over 4} q_{i}^{4} {\Bigr)}
+{1\over 4} \sum_{i{\;}{\ge}{\;}j=0}^{N-1} c_{ij}{\Bigl(}q_{i}
-q_{j}{\Bigr)}^{4}
{\;}{\equiv}{\;}
\sum_{i=0}^{N-1}H_{i} + \sum_{i{\;}{\ge}{\;}j=0}^{N-1} H_{ij},
$$
with $q_{N}{\;}{\equiv}{\;}q_{0}$,
and view the system as a sum of $N$ individual one-degree-of-freedom 
Hamiltonians $H_{i}$ plus an interaction Hamiltonian $H_{int}=\sum H_{ij}$.
To explore how the dynamics depends on both the range of the interaction
and the number of degrees of freedom that are coupled directly, one can then
allow for three different types of couplings:
\begin{enumerate}
\item{a model with nearest neighbor couplings, with $c_{ij}$ nonzero only 
for $|i-j|=1$, and with all nonzero $c_{ij}$'s assuming the same value;}
\item{a maximally connected model, with all pairs coupled identically;
and}
\item{another long-range model, where the coupling constant decreases linearly
from a fixed value $c$ for nearest neighhbors to zero strength for oscillators
separated by $N/2$ `spaces.'}
\end{enumerate}
Unless otherwise specified, the computations described here assumed a 
coupling strength $c_{i,i+1}=1/3$.

As a highly special, albeit interesting, initial condition, suppose that,
at $t=0$, all the system energy is deposited into a single degree of freedom,
setting the kinetic energy for one of the oscillators, say oscillator $0$,
equal to the total system energy $E$. The obvious question then is 
how long it takes before a sizeable fraction of the energy is transferred
to the other oscillators and/or the interaction Hamiltonian.

Consider first models with nearest neighbor couplings. In this
case, one discovers that a significant fraction of the total energy
can remain trapped in a single degree of freedom for times as long as
${\sim}{\;}10^{2}-10^3$ periods of uncoupled oscillations. This is, 
{\em e.g.,} illustrated in Figs.~1 and 2, which exhibit the 
distribution of kinetic energies amongst the different oscillators as a
function of time for four different energies. Here, as in Figs.~3 - 5, the 
kinetic energies have been smoothed from raw data recorded at intervals 
${\delta}t=0.1$ by a box-car averaging over 400 adjacent points.
The initial localization involves of course the oscillator in which the 
energy was originally deposited. Eventually, however, different degrees 
of freedom emerge as the location of this energy localization, the 
transitions occuring quite abruptly and with no obvious correlation 
between the locations of the localization prior to and following the 
transitions. The sojourn times also appear to be random, except that 
for higher energies and/or weaker coupling, the intervals between
the successive transitions become shorter overall. For example, in the 
long-range model with linear decrease in coupling strength, reducing 
$c$ from $1/3$ to $1/9$ can decrease the typical duration of these intervals 
by as much as a 
factor of five. For very high energies, 
the localized configuration breaks down rapidly, never to appear
again (see, {\em e.g.,} Fig. 2, bottom panel).

The obvious question is how this conclusion is altered for the case of 
longer range interactions. In particular,
does increasing the range of the coupling make the energy `disperse' more 
quickly? Here the answer is a resounding: {\em no}! Far from providing new 
channels for energy transfer between the degrees of freedom, {\it long-range 
couplings make this energy localization, or trapping, even more robust.} For
every value of energy that was explored 
-- $0.02{\;}{\le}{\;}E{\;}{\le}{\;}327.68$ --, localization was more robust
for the models that coupled together all the degrees of freedom than for
the nearest neighbor model, with the maximally connected model allowing 
trapping for the longest time. 
The degree to which a long range coupling enhances trapping can be inferred
from Fig.~3, which exhibits the distribution of kinetic energies for the
different oscillators for the three different models, in each case allowing
for an energy $E=327.68$.
Also evident from Figs.~1 - 3 is the fact that, for all three models, the 
`trapping time'
decreases with increasing energy, a relative sharp decrease being observed
for energies above $E_{*}{\;}{\sim}{\;}5-20$, the value of which depends on 
the model.

However, as illustrated by Fig.~4, localization does not emerge for 
`more generic' initial conditions 
with energy distributed randomly amongst all the oscillators, so one might 
perhaps conjecture that this localization is a fluke reflecting a very 
nongeneric initial condition. It is therefore important to determine the 
stability of these localized configurations. In particular, one needs to
determine whether, for initial conditions `less singular' than $E_0 = E$, 
energy trapping persists and, if so, for how long.

These issues were addressed by considering alternative sets of initial
conditions, where the initial kinetic energy of the zeroeth oscillator was
assigned smaller fractions of the energy, $E_{0}=0.85E$ and $E_{0}=0.7E$,
and the remaining energy was apportioned at random in the other degrees of
freedom. Putting energy into the other degrees
of freedom does indeed tend to make localization less robust. However, even
for $E_{0}=0.7E$ one can see distinct localization patterns that persist over
hundreds of natural oscillation periods. Moreover, as before, stronger and/or 
longer range couplings mean that the localization persists for longer times.
Decreasing the fraction of the total energy placed into the `trap' results
in shorter trapping times. This behavior is illustrated in Fig.~5, 
where different coupling types and values of $E$ are chosen so as to 
demonstrate the competing effects of longer range coupling and higher energy. 
In all three cases, the initial condition is one with 70\% of the total 
energy deposited in oscillator 0.

The middle
panel of this Figure also illustrates another interesting point: It is 
possible
for most of the energy localized initially in a single oscillator to be
deposited in two other oscillators, where it remains localized for a 
comparatively long time. Alternatively, as illustrated in the bottom panel of
Fig.~1, an initial state which would appear to have become largely 
delocalized 
can `relocalize' with most of the energy concentrated in a different 
oscillator.

Since the efficacy of mixing is well known to differ for regular and chaotic
dynamics, it is also natural to ask whether the comparatively abrupt 
transition
from robust to less stable trapping with increasing energy $E$ 
correlates with a transition from (near-)regular to (more strongly) chaotic
dynamics. One useful diagnostic in addressing such a transition is provided
by examining the `complexity' of the time series associated with some phase
space variable, such as the position, momentum, or energy of one of the
oscillators. One can, {\em e.g.,} determine the number of frequencies 
required,
on average, to capture some significant fraction, (say) $95\%$, of the total 
power in the Fourier spectrum associated with some time series. 

Such an analysis reveals  that, for lower energies, the time series are very 
close to periodic whereas, for higher energies, any quasi-periodic 
approximation 
requires an enormous number of frequencies. There is, moreover, a strong
correlation between the magnitude of the time series complexity and the
rate of delocalization. The transition from very slow to considerably 
more rapid delocalization and the transition from very long-lived to 
considerably less long-lived trapping are comparably abrupt. And
the complexity for the shorter range models, where localization is less 
robust, tends to be somewhat larger than for the time series of the longer 
range
models. Finally, it is evident that, at least while energy remains trapped
in one oscillator, the time series for the coordinate or momentum 
corresponding
to that oscillator 
is typically much less complex than the the time series
for the other degrees of freedom. Alternatively, at least for the case of the
maximally connected model, the oscillator in the chain directly opposite from 
the oscillator in which the energy is localized tends to have an especially
complex time series. 

Examples of this behavior are exhibited in Fig.~6,
which derive from orbital data recorded at intervals ${\delta}t=0.1$ for a
total time $t=3200$. In each case, $n(0.95)$ represents the fraction of 
frequencies
required to capture 95\% of the power in a time series for phase space 
coordinates $q_{i}$ and $p_{i}$. The bottom curve exhibits data for the 
oscillator into which all the energy was originally deposited, the diamonds
representing mean values obtained by computing complexities individually
for $q_{0}$ and $p_{0}$ and then averaging. The top curve exhibits data for
the oscillator separated by $N/2$ `spaces'. The middle curve was derived by
computing complexities analogously for all $16$ oscillators and then
constructing an average over the oscillators.
 
Presuming that the flow is ergodic, one would expect an eventual 
evolution towards a `well mixed' state; and, for that reason, it
is natural to determine the extent to which there is an asymptotic approach 
towards equipartitioning of energy at late times. For the case of random 
initial conditions, there are in fact clear indications of such an approach, 
at least in a time-averaged sense, although the time scale for this approach
can be very long, $>10^{3}$ orbital periods. Obviously, though, for 
localized 
initial states there can be no such approach as long as trapping persists. 
However, even here there are some indications for an eventual approach 
towards 
equipartitioning for higher energies, weaker couplings, and/or shorter range 
interactions. Examples of an approach towards equipartitioning, or lack 
thereof, are provided in Fig.~7, which tracks the time-averaged kinetic
energies ({\em cf.} Batt 1987)
\begin{displaymath}
{\langle}K(t){\rangle}={1\over n}\sum_{i=1}^{n}K(t_{i}),
\end{displaymath}
with $t_{1}=0$ and $t_{n}=t$, of several different oscillators in three 
representative integrations. 

A number of interesting questions still remain to be addressed. For each of
the different types of coupling, is there, {\em e.g.,} a threshold value 
of energy or coupling strength signaling a transition from regularity to
chaos and a comparatively rapid breakdown of localization? How does the
behavior observed for $N=16$ vary as the number of degrees of freedom 
increases? And, perhaps most importantly, is this localization unique to
one-dimensional chains? Will a similar localization persist on a 
two-dimensional lattice configured as a torus? 
Work on these questions is currently underway.

In any event, the numerical experiments performed hitherto yield three
unambiguous conclusions: (1) {\em FPU}-type 
systems with both nearest-neighbor and longer range couplings 
admit states in which most of the energy remains trapped in a single degree of
freedom for relatively long times. (2) The stronger the coupling in terms 
of range, connectance, or size of the coupling constant $c$, the more robust 
is this energy
trapping. (3) This trapping is more robust at lower energies, where the
dynamics seems more nearly regular.
\vskip .2in
\par\noindent HEK was supported in part by NSF-AST-0070809.
\vskip .2in
\par\noindent {\bf References}
\vskip .1in
\par\noindent
Batt, J. 1987, Transport Theory Stat. Phys. 16, 763
\par\noindent
Ford, J. 1992, Phys. Repts. 213, 271 
\par\noindent
Froeschl\'e, C. 1978, Phys. Rev. A 18, 277 
\vfill\eject
\pagestyle{empty}
\begin{figure}[t]
\centering
\centerline{
        \hskip -3.0in
        \epsfxsize=20cm
        \epsffile{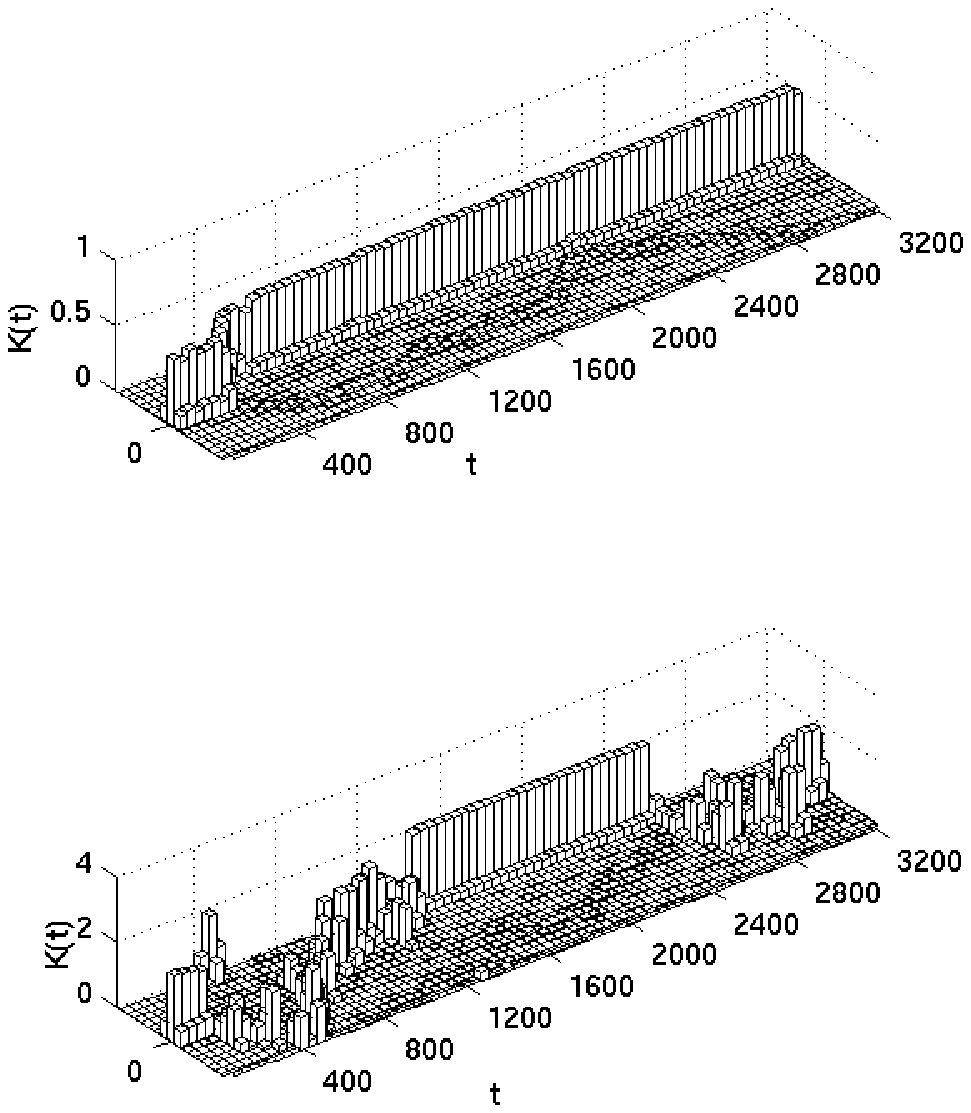}
           }
        \begin{minipage}{10cm}
        \end{minipage}
        \vskip -3.2in\hskip -0.1in
\caption{Energy trapping in the nearest neighbor model. (top) Distribution 
of kinetic energies as a function of time for an initial condition with all 
the energy $E=1.28$ originally in oscillator 0.
(bottom) The same for $E=5.12$. 
}
\vspace{0.1in}
\end{figure}

\begin{figure}[t]
\centering
\centerline{
        \hskip -3.0in
        \epsfxsize=20cm
        \epsffile{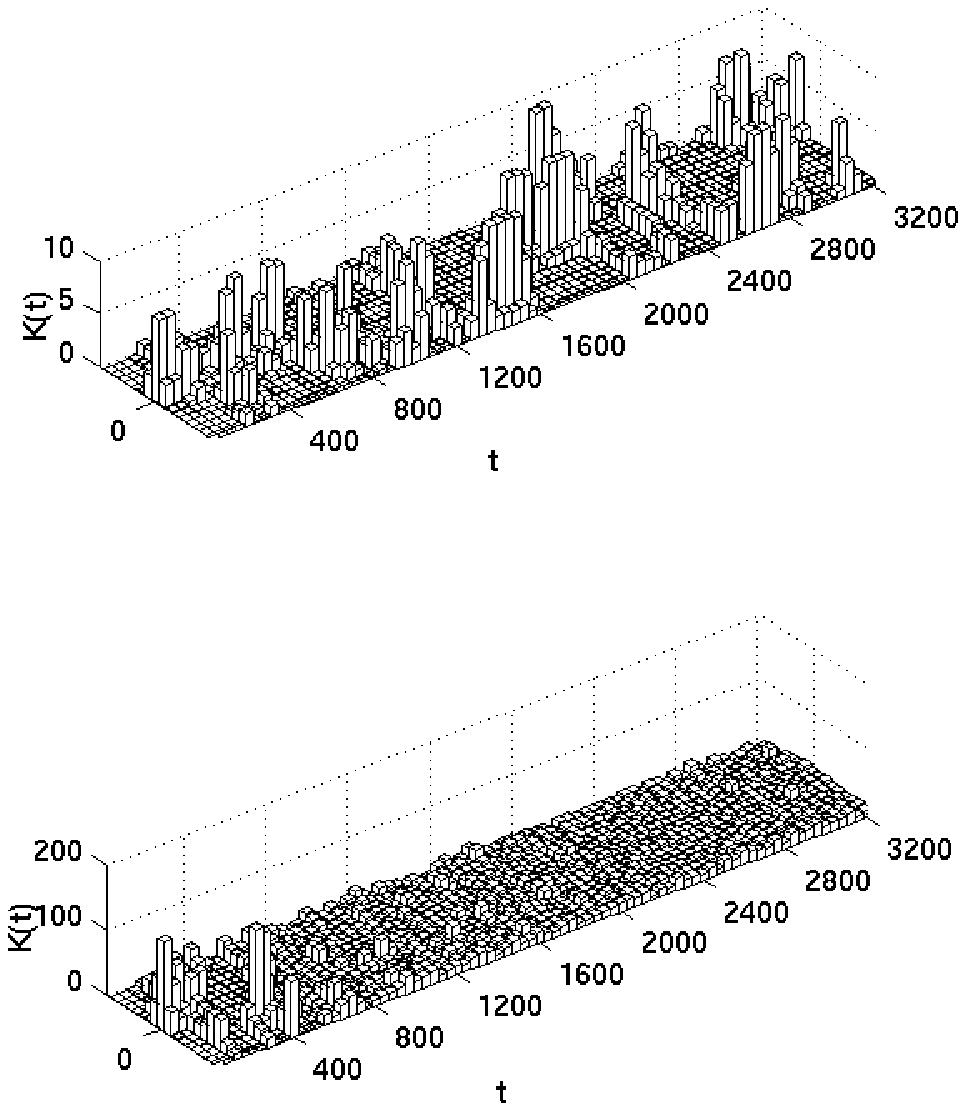}
           }
        \begin{minipage}{10cm}
        \end{minipage}
        \vskip -3.2in\hskip -0.1in
\caption{Energy trapping in the nearest neighbor model. (top) Distribution 
of kinetic energies as a function of time for an initial condition with all 
the energy $E=20.48$ originally in oscillator 0. 
(bottom) The same for $E=327.68$. 
}
\vspace{0.1in}
\end{figure}

\begin{figure}[t]
\centering
\centerline{
        \hskip -4.0in
        \epsfxsize=24cm
        \epsffile{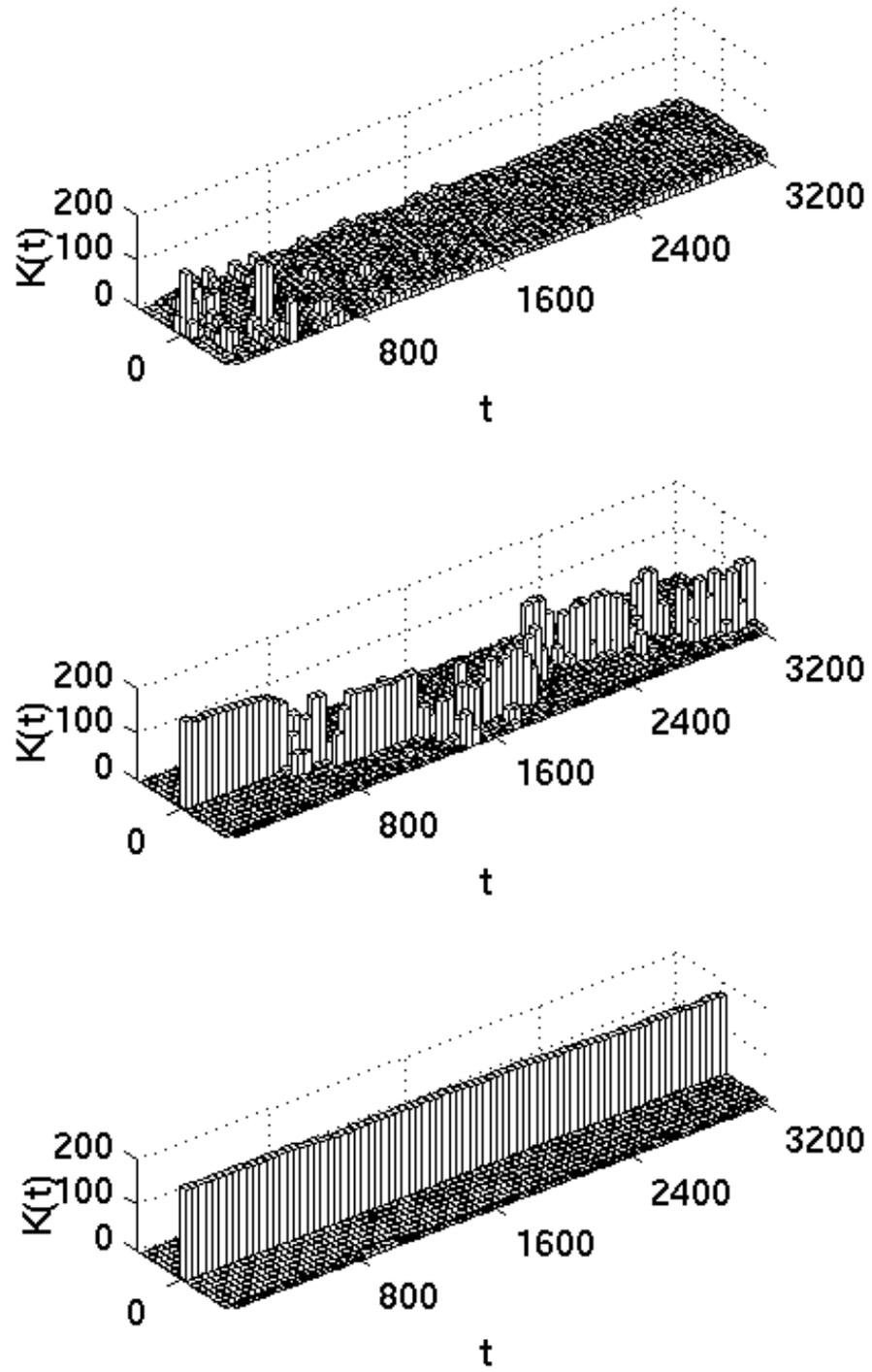}
           }
        \begin{minipage}{10cm}
        \end{minipage}
        \vskip -4.8in\hskip -0.1in
\caption{Distributions of kinetic energies as a function of time for
an initial condition with the energy $E=327.68$ all originally 
in oscillator 0. (top) The nearest neighbor model. (middle) The long range
model with coupling decreasing linearly.
(bottom) The maximally connected model.
}
\vspace{0.1in}
\end{figure}

\begin{figure}[t]
\centering
\centerline{
        \hskip -3.0in
        \epsfxsize=20cm
        \epsffile{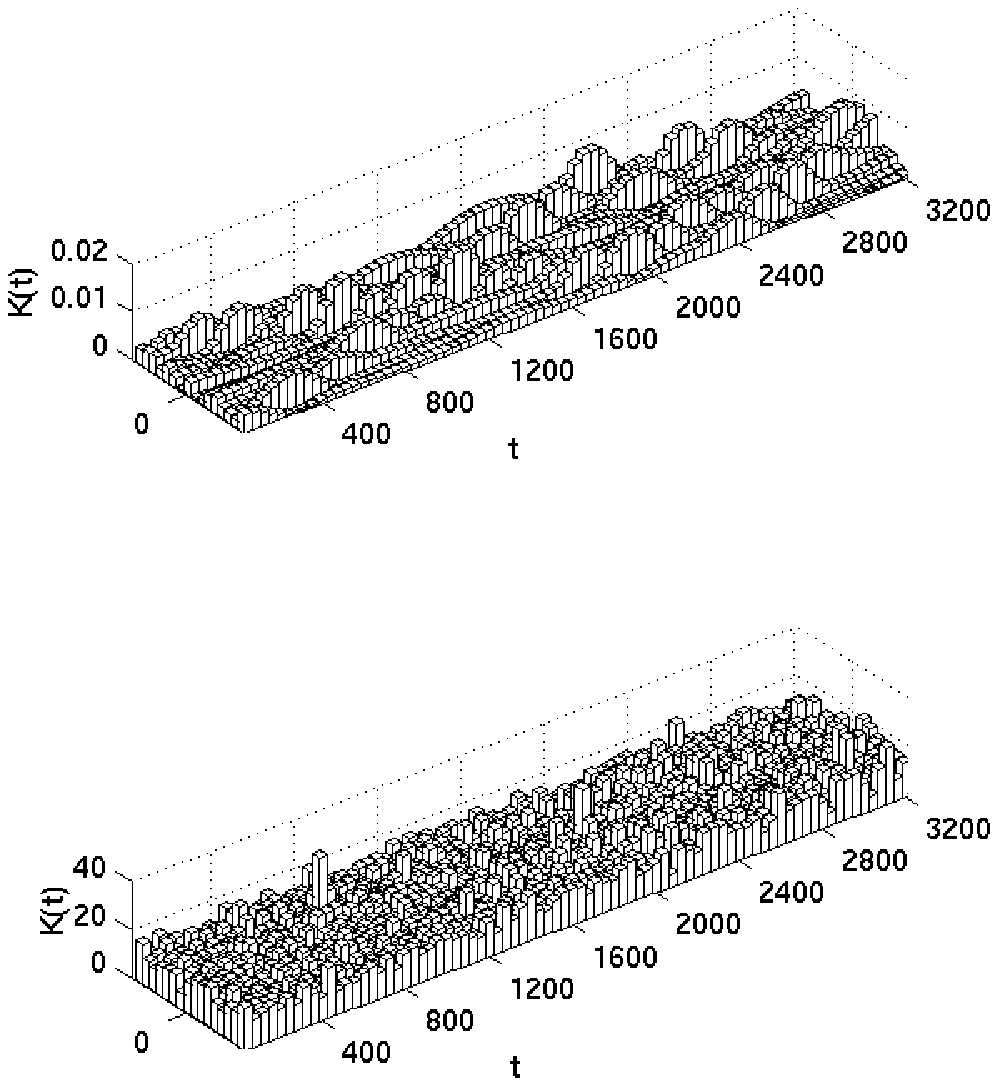}
           }
        \begin{minipage}{10cm}
        \end{minipage}
        \vskip -3.2in\hskip -0.1in
\caption{(top) Distribution of kinetic energies as a function of time for
a random initial condition with $E=0.08$ evolved in the nearest neighbor 
model. (bottom) The same for $E=327.68$.
}
\vspace{0.1in}
\end{figure}

\begin{figure}[t]
\centering
\centerline{
        \hskip -4.0in
        \epsfxsize=24cm
        \epsffile{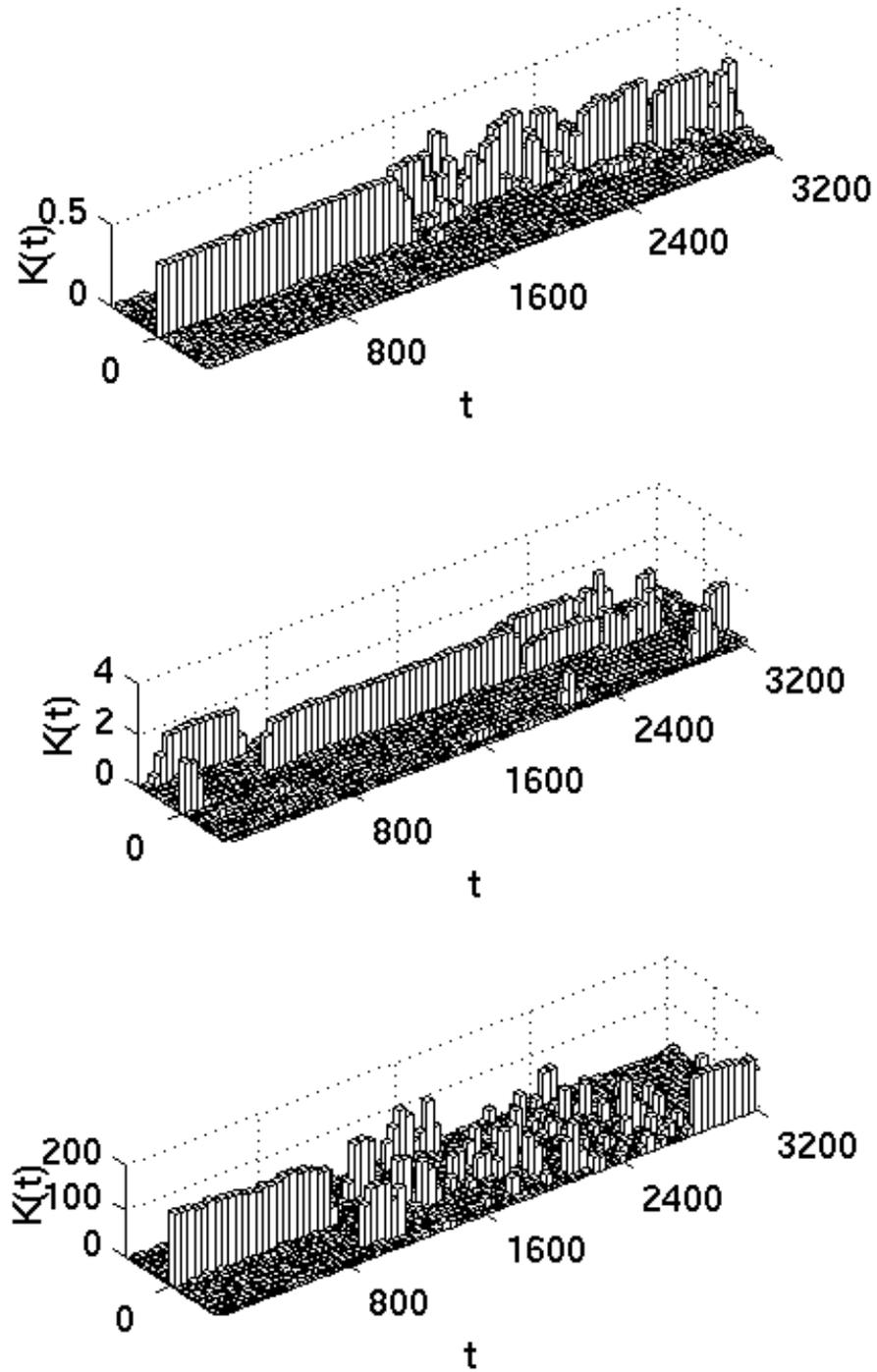}
           }
        \begin{minipage}{10cm}
        \end{minipage}
        \vskip -4.8in\hskip -0.1in
\caption{(top) Distribution of kinetic energies as a function of time for
an initial condition with $E_{0}=0.70E$ evolved in the linearly decreasing 
long-range coupling model with total energy $E=1.28$. The remaining energy
is initially distributed randomly amongst the other degrees of freedom.
(middle) The same for an initial condition with $E=5.12$ evolved in the
maximally connected model. (bottom) An initial condition with $E=327.68$
evolved in the maximally connected model.
}
\vspace{0.1in}
\end{figure}

\begin{figure}[t]
\centering
\centerline{
        \epsfxsize=10cm
        \epsffile{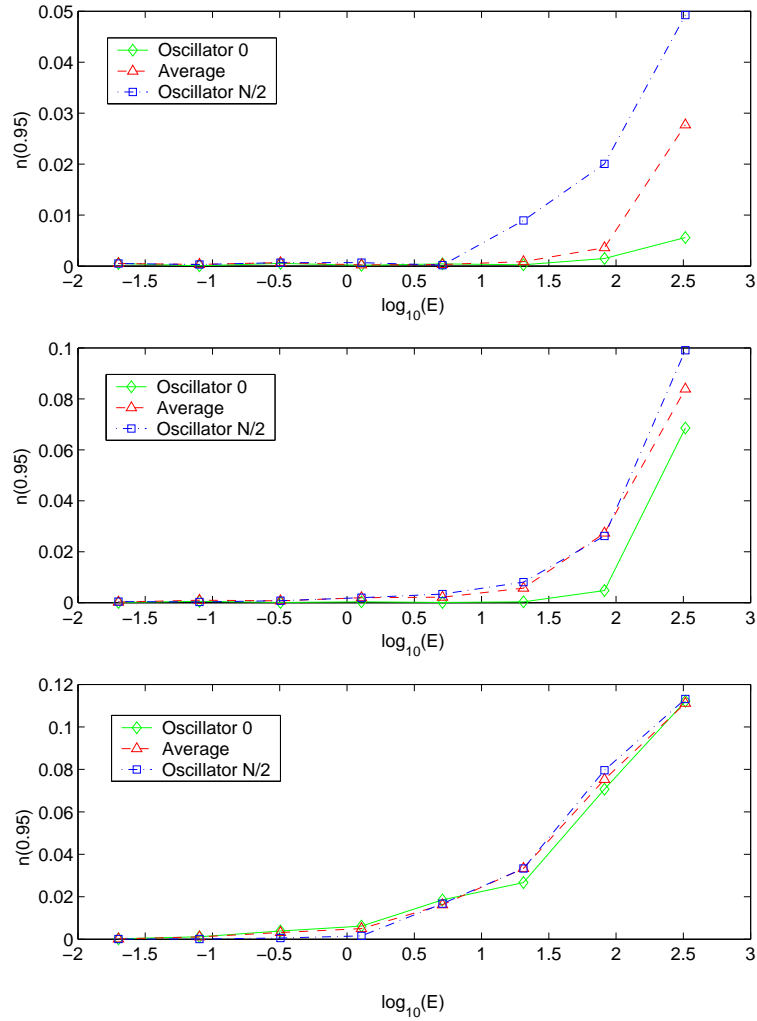}
           }
        \begin{minipage}{10cm}
        \end{minipage}
        \vskip -0.2in\hskip -0.1in
\caption{(top) Orbital complexities as a function of energy $E$, computed for 
different degrees of freedom in the maximally connected model from initial
conditions in which all the energy is initially localized in a single degree
of freedom. The bottom curve is for the oscillator which receives the initial
energy, the top curve is for the oscillator separated by $N/2$ `spaces', and 
the middle curve represents an average over all 16 oscillators.
(middle) The same for the linear long range model.
(bottom) The same for the nearest neighbor model.
}
\vspace{0.1in}
\end{figure}

\begin{figure}[t]
\centering
\centerline{
        \epsfxsize=14cm
        \epsffile{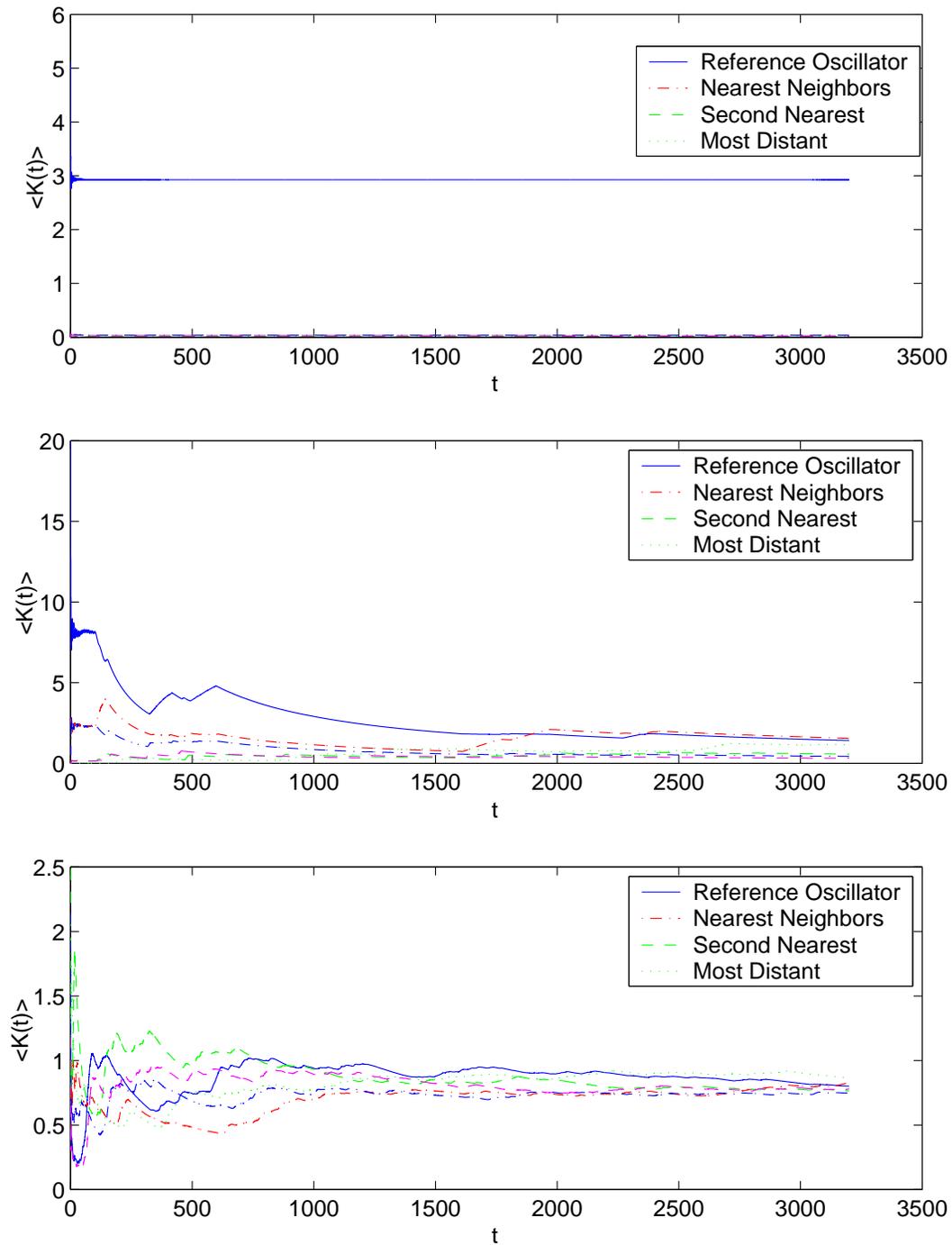}
           }
        \begin{minipage}{10cm}
        \end{minipage}
        \vskip -0.2in\hskip -0.1in
\caption{(top) Time-averaged kinetic energies ${\langle}K(t){\rangle}$ 
for oscillators in the model 
with long-range couplings that decrease linearly, evolved from an initial
state in which all the energy $E=5.12$ is deposited in one oscillator.
(middle) The same for the model with nearest neighbor couplings, now 
constructed for an initial state with $E=20.48$.
(bottom) The same for a random initial condition with $E=20.48$ evolved in 
the model with maximal coupling.
}
\vspace{0.1in}
\end{figure}
\end{document}